\documentstyle[12pt]{article}
\begin{document}
\begin{center} {\large\bf SOME ASPECTS OF NUCLEAR DEEP
INELASTIC SCATTERING} \end{center}
\vskip 1em
\begin{center} {\large Felix M. Lev} \end{center}
\vskip 1em
\begin{center} {\it Laboratory of Nuclear
Problems, Joint Institute for Nuclear Research, Dubna, Moscow region
141980 Russia (E-mail:  lev@nusun.jinr.dubna.su)} \end{center}
\vskip 1em
\begin{abstract}
Nuclear deep inelastic scattering is considered in the framework of a
model in which the current operator explicitly satisfies Poincare
invariance and current conservation. The results considerably differ
from the standard ones at small values of the Bjorken variable $x$. In
particular, it is impossible to extract the neutron structure functions
from the deuteron data at $x
\hbox{{\normalsize\,\,%
\lower -.6ex \hbox{$<$}\kern -.75em \lower .5ex \hbox{$\sim$}\,\,}}
0.01$ and we predict that the behavior of the
deuteron structure functions at low $x$ and large momentum transfer
considerably differs from the behavior of the nucleon structure functions
at such conditions. We also argue that for heavier nuclei the effect of
the final state interaction is important even in the Bjorken limit.
\end{abstract}

\section{Introduction}
\label{S1}

  There exists a vast literature devoted to the problem of describing
nuclear deep inelastic scattering (DIS). After the discovery of the
original EMC effect \cite{EMC} the central point of the extensive
discussion was whether this effect can be explained in the framework
of conventional nuclear physics. The point of view advocated by many
authors is that nuclear DIS can be described with a good accuracy if
only the nucleon degrees of freedom in the nucleus wave function are
taken into account but relativity of the internal motion of nucleons
in the nucleus is rather important.

 The first calculations in such a framework were carried out in Refs.
\cite{Ak} and others; as pointed out in Ref. \cite{FS}, the important
role in these calculations plays the "flux factor". The status of
nuclear DIS at the end of 80th is discussed in detail in Ref.
\cite{Bick}. Since that time the convolution formula for nuclear DIS
was rederived by several authors using different considerations
(see Refs. \cite{Ciofi} and others). However, as argued in Ref.
\cite{MelKul}, the convolution formula can be obtained only if some
simplifying assumptions are made, and in particular the off-shellness
of the struck nucleon is neglected.

  Let $P'$ be the 4-momentum of the initial nucleus and $q$ be the
four-momentum of the virtual photon absorbed by this nucleus (for
definiteness we shall speak about electromagnetic interactions but the
same is valid for weak ones). We shall consider nuclear DIS only in the
Bjorken limit when $|q^2|\gg m^2$, where $m$ is the nucleon mass, but
$x=-q^2/2(P'q)$ is not too close to 0 or 1.

 The idea adopted in practically all works on nuclear DIS is that, by
analogy with the well-known nonrelativistic calculations in atomic and
nuclear physics, nuclear DIS in the Bjorken limit can be described in
the impulse approximation (IA). By definition, the IA implies that the
electromagnetic or weak current operator (CO) of the system under
consideration is a sum of the CO's for the constituents comprising this
system. Therefore the IA in nuclear DIS implies that the nucleus CO is
a sum of the CO's for the nucleons comprising the given nucleus, and
the IA in nucleon DIS implies that the CO is a sum of the quark CO's.

 In the nonrelativistic case we always work in such a representation
of the generators of the Galilei group when only the Hamiltonian is
interaction dependent while all other nine generators are free.
This choice of the representation is reasonable since time flows
equally in all reference frames. However, as shown by Dirac \cite{Dir},
in the relativistic case at least three of ten representation generators
of the Poincare group are interaction dependent. If we wish to use the
IA in the relativistic case then the question immediately arises in
which representation of the Poincare group the CO can be taken as a
sum of the constituent CO's. Different representations of the Poincare
group describe the same physics if these representations are unitarily
equivalent, but the unitary operators realizing the equivalence are
(generally speaking) interaction dependent. Therefore if the CO is a
sum of the constituent CO's in some representation, then (as pointed
out by several authors) this property does not take place in other
representations.

 The system CO ${\hat J}^{\mu}(x)$, where $\mu=0,1,2,3$ and $x$ is a
point in Minkowski space, should satisfy the following necessary
conditions.

 First, let ${\hat U}(a)=exp(\imath {\hat P}_{\mu}a^{\mu})$ be the
representation operator corresponding to the
displacement of the origin in spacetime translation of Minkowski
space by the 4-vector $a$. Here ${\hat P}=({\hat P}^0,{\hat {\bf
P}})$ is the operator of the 4-momentum, ${\hat P}^0$ is the
Hamiltonian, and ${\hat {\bf P}}$ is the operator of ordinary
momentum. Let also ${\hat U}(l)$ be the representation operator
corresponding to $l\in SL(2,C)$. Then ${\hat  J}^{\mu}(x)$ must be
the relativistic vector operator such that
\begin{equation}
{\hat U}(a)^{-1}{\hat  J}^{\mu}(x){\hat U}(a)=
{\hat J}^{\mu}(x-a)
\label{1}
\end{equation}
\begin{equation}
{\hat U}(l)^{-1}{\hat J}^{\mu}(x){\hat U}(l)=L(l)^{\mu}_{\nu}
{\hat J}^{\nu}(L(l)^{-1}x)
\label{2}
\end{equation}
where $L(l)$ is the element of the Lorentz group corresponding to $l$
and a sum over repeated indices $\mu,\nu=0,1,2,3$ is assumed.

 Second, if ${\hat J}^{\mu}(x)$ satisfies the continuity equation
$\partial {\hat J}^{\mu}(x)/\partial x^{\mu}=0$ then, as follows from
Eq. (\ref{1}), this equation can be written in the form
\begin{equation}
[{\hat  J}^{\mu}(x),{\hat P}_{\mu}]=0
\label{3}
\end{equation}

 Finally, the operator ${\hat J}^{\mu}(x)$ should also satisfy the
cluster separability condition. Briefly speaking, this condition
implies that if the interaction between any subsystems
$\alpha_1,...\alpha_n$ comprising the system under consideration is
turned off then ${\hat J}^{\mu}(x)$ must become a sum of the current
operators ${\hat J}_{\alpha_i}^{\mu}(x)$ for the subsystems.

 As pointed out by Dirac \cite{Dir}, any physical system can be
described in different forms of relativistic dynamics. Let
${\hat M}^{\mu\nu}$ (${\hat M}^{\mu\nu}=-{\hat M}^{\nu\mu}$) be the
generators of the Lorentz group. We use $P$ and $M^{\mu\nu}$ to
denote the 4-momentum operator and the generators of the Lorentz
group in the case when all interactions are turned off. By definition,
the description in the point form implies that the operators
${\hat U}(l)$ are the same as for noninteracting particles, i.e.
${\hat U}(l)=U(l)$ and ${\hat M}^{\mu\nu}=M^{\mu\nu}$, and thus
interaction terms can be present only in the 4-momentum operators
${\hat P}$ (i.e. in the general case ${\hat P}^{\mu}\neq P^{\mu}$ for
all $\mu$). The description in the instant form implies that the
operators of ordinary momentum and angular momentum do not depend on
interactions, i.e. ${\hat {\bf P}}={\bf P}$, ${\hat {\bf M}}={\bf M}$
$({\hat {\bf M}}=({\hat M}^{23},{\hat M}^{31},{\hat M}^{12}))$ and
therefore interactions may be present only in ${\hat P}^0$ and the
generators of the Lorentz boosts ${\hat {\bf N}}=({\hat M}^{01},
{\hat M}^{02},{\hat M}^{03})$. In the front form with the marked $z$
axis we introduce the + and - components of the 4-vectors as $p^+=
(p^0+p^z)/\sqrt{2}$, $p^-=(p^0-p^z)/\sqrt{2}$. Then we require that
the operators ${\hat P}^+,{\hat P}^j,{\hat M}^{12},{\hat M}^{+-},
{\hat M}^{+j}$ $(j=1,2)$ are the same as the corresponding free
operators and therefore interaction terms may be present only in the
operators ${\hat M}^{-j}$ and ${\hat P}^-$.

 As follows from the above formulas and definitions there is no form of
dynamics in which the IA is valid. Let us note however that as follows
from Eq. (\ref{1})
\begin{equation}
{\hat J}^{\mu}(x)=exp(\imath {\hat P}x){\hat J}^{\mu}(0)
exp(-\imath {\hat P}x)
\label{4}
\end{equation}
Therefore, if the operator ${\hat P}$ is known, the problem of
constructing ${\hat J}^{\mu}(x)$ can be reduced to that of
constructing the operator ${\hat J}^{\mu}(0)$ with the correct
properties.

 In turn, as follows from Eq. (\ref{2}), Lorentz invariance of the
CO implies
\begin{equation}
[{\hat M}^{\mu\nu},
{\hat J}^{\rho}(0)]= -\imath({\eta}^{\mu\rho}{\hat J}^{\nu}(0)-
{\eta}^{\nu\rho}{\hat J}^{\mu}(0))
\label{5}
\end{equation}
where ${\eta}^{\mu\nu}$ is the metric tensor in Minkowski space.

  While the IA for the operator ${\hat J}^{\mu}(x)$ is incompatible
with Eqs. (\ref{1}) and (\ref{2}), the question arises whether this
approximation can be valid for the operator ${\hat J}^{\mu}(0)$.
 By definition, the operator ${\hat J}^{\mu}(0)$ in the IA is given by
\begin{equation}
{\hat J}^{\mu}(0)=J^{\mu}(0)=\sum_{i=1}^{N}J_i^{\mu}(0)
\label{6}
\end{equation}
where $J_i^{\mu}(0)$ is the CO for the i-th nucleon at $x=0$. In this case
cluster separability is satisfied automatically, though in the general
case this condition is rather restrictive \cite{KlPol,lev}.

 As noted above, in the instant and front forms some of the operators
${\hat M}^{\mu\nu}$ are interaction dependent. Since the free operators
$M^{\mu\nu}$ satisfy Eq. (\ref{5}) if ${\hat J}^{\mu}(0)$ is given by Eq.
(\ref{6}), Eqs. (\ref{5}) and (\ref{6}) will take place in these forms
only if the interaction terms in ${\hat M}^{\mu\nu}$ commute with
$J^{\rho}(0)$. However neither in systems with a fixed number of particles,
nor in quantum field theory there is no reason for such a commutation
\cite{lev,qft}. Let us also note that, as shown by many authors (see, for
example, Ref. \cite{Web}), the parton model is a consequence of the IA in
the front form, and therefore Eq. (\ref{5}) is not satisfied in this case
(see Ref. \cite{hep} for a detailed discussion).
However in the point form the operators
${\hat M}^{\mu\nu}$ are free and therefore in the point form the IA for
the operator ${\hat J}^{\mu}(0)$ is compatible with Lorentz invariance.

 Let us now consider the constraints imposed on ${\hat J}^{\mu}(0)$ by
the continuity equation (\ref{3}). Let $m_A$ be the mass of the initial
state $|i\rangle$ and $G'$ be its four-velocity such that $P'=m_AG'$.
Consider the transition of the nucleus to the final state $|f\rangle$
with the mass $M"$ and four-velocity $G"$ such that the total momentum
in the final state is equal to $P"=M"G"$.
Since the Lorentz boost operators in the point form are free, there is
no problem in boosting the wave function (WF) from one reference frame
to another. As shown in Ref. \cite{lev}, the current operator is fully
defined by its matrix elements in the reference frame where
${\bf G}"+{\bf G}'=0$. The coordinate axes in this frame can be
chosen in such a way that ${\bf G}_{\bot}"={\bf G}'_{\bot}=0$
(where we use the
subscript $\bot$ to denote the projection of the three-dimensional
vector onto the plane $xy$) and the $z$ axis is directed along the
momentum of the initial nucleus. Then $G^{"0}=G^{'0}$, $G^{"z}=-G^{'z}$
and Eq. (\ref{3}) for $x=0$ can be written in the form
\begin{equation}
(M"G^{'-}-m_AG^{'+})\langle f|{\hat J}^-(0)|i\rangle =
-(M"G^{'+}-m_AG^{'-})\langle f|{\hat J}^+(0)|i\rangle
\label{7}
\end{equation}

 Let us suppose that the four-velocity operator ${\hat G}$ does not
contain interaction terms. Then the first consequence of Eq. (\ref{7})
is that the $\bot$ components of ${\hat J}^{\mu}(0)$ are not constrained by
the continuity equation, and therefore we can take the IA for these
components. Since $(M")^2=(P'+q)^2$ and $P'+q=M"G"$, it is easy to show
that
\begin{equation}
(M")^2=m_A^2-\frac{q^2(1-x)}{x}
\label{8}
\end{equation}
and in the reference frame under consideration
\begin{equation}
(G^{'z})^2=\frac{|q^2|^{1/2}}{4m_A[x(1-x)]^{1/2}},\quad
q^0=(M"-m_A)G^{'0},\quad {\bf q}=-(M"+m_A){\bf G}'
\label{9}
\end{equation}
Therefore, as follows from Eqs. (\ref{7}-\ref{9}),
\begin{equation}
\langle f|{\hat J}^+(0)|i\rangle =\frac{m_Ax}{M"}\langle
f|{\hat J}^-(0)|i\rangle
\label{10}
\end{equation}
We conclude that in the reference frame under consideration the matrix
elements of the operator ${\hat J}^+(0)$ in the Bjorken limit are
negligible in comparison with the matrix elements of the operator
${\hat J}^-(0)$. Therefore a possible prescription is to take
${\hat J}^-(0)$ in the IA and then
the matrix
elements of the operator ${\hat J}^+(0)$ can be defined (if necessary)
from Eq. (\ref{10}). However this procedure is not unique.

 Many authors realize the idea of the IA not on the operator level but
in the framework of the covariant approach based on Feynman diagrams.
Namely the IA is associated with the disconnected (or one-leg)
diagrams, i.e. the diagrams in which the virtual photon interacts with
only one nucleon while the other nucleons are spectators. The problem
whether the contribution of the connected diagrams is indeed negligible
in the Bjorken limit was investigated by several authors (see, for
example, Refs. \cite{LanPol,EFP} and references therein). The main
motivation is that if ${\tilde p}$ is the four-momentum of the struck
constituent in the disconnected diagram then $|{\tilde p}^2|\ll |q^2|$
while for all particles in the connected diagrams the off-shellness is of
order $|q^2|$. However, even if some connected diagram can be reliably
estimated, the problem exists what is the sum of all such diagrams.
The above considerations show that the very notion of the IA is not
covariant since even for the operator ${\hat J}^{\mu}(0)$ the IA can be
consistent only at some special choice of the representation of the
Poincare group.

 In view of the above discussion it seems reasonable to consider nuclear
DIS in the Bjorken limit if the nucleus dynamics is described in the
point form and the operator ${\hat J}^{\mu}(0)$ is given by Eq. (\ref{6}).
In Sec. \ref{S2} we describe the nucleus WF in the point form, and in
Sec. \ref{S3} standard formulas for DIS are written down. The main
result of the paper is given in Sec. \ref{S4} where we derive in detail
the relations between the deuteron and nucleon structure functions.
In Sec. \ref{S5} the case of heavier nuclei is briefly considered, and
Sec. \ref{S6} is discussion.

\begin{sloppypar}
\section{Nucleus wave function in the point form of dynamics}
\label{S2}
\end{sloppypar}

 In local quantum field theory the number of particles is not a
conserving physical quantity, and one might think that if the nucleus is
described relativistically then it cannot be considered as a system of a
fixed number of nucleons. However there are grounds to believe that in few
nucleon systems at low energies the main part of the relativistic
corrections are due to relativistic kinematics rather than the excitation
of new degrees of freedom (see, for example, the discussion in Refs.
\cite{Kon,KeiPol,riv}). Therefore a good approximation to the description
of relativistic effects is relativistic quantum mechanics, i.e.
the relativistic theory of systems with a fixed number of particles.
In this theory it is possible to explicitly construct all the
representation generators of the Poincare group in such a way that
cluster separability is satisfied \cite{sok,CP,Mutze,lev1}.

 A rather strange feature of the present approaches to DIS is that while
nucleon DIS is usually considered in the infinite momentum frame (IMF),
nucleus DIS is usually considered in the reference frame where the
initial nucleus is at rest. The matter is that in usual approaches the
WF of a composite system essentially depends on the
reference frame (this is clear, for example, in the instant form where
the Lorentz boost generators are interaction dependent). Therefore it
is desirable to work in the reference frame in which the WF has the
simplest form. It is believed that such a frame for the nucleon is IMF
(where the nucleon is a system of almost free partons) while for the
nucleus this is the rest frame where the nucleus can be approximately
described as a nonrelativistic system of nucleons
(see the discussion in Ref. \cite{Br}). However, as we already noted,
in the point form there is no problem in boosting the WF from one
reference frame to another.

 The Hilbert space for the system of $A$ (free or interacting) nucleons
is the space of functions $\varphi({\bf p}_{1\bot},p_1^+,\sigma_1,...
{\bf p}_{A\bot},p_A^+,\sigma_A)$ where $p_i$ and $\sigma_i=\pm 1/2$
$(i=1,...A)$ are the four-momentum and the spin projection on the $z$
axis for the i-th nucleon,
\begin{equation}
\sum_{\sigma_1...\sigma_A}\int\nolimits |\varphi({\bf p}_{1\bot},p_1^+,
\sigma_1,...{\bf p}_{A\bot},p_A^+,\sigma_A)|^2 \prod_{i=1}^{A}
d\rho({\bf p}_{i\bot},p_i^+)\quad< \infty
\label{11}
\end{equation}
and
\begin{equation}
d\rho({\bf p}_{\bot},p^+)= \frac{d^2{\bf p}_{\bot}dp^+}
{2(2\pi)^3p^+}
\label{12}
\end{equation}

 We define $P=p_1+...+p_A$, $M_0=|P|\equiv (P^2)^{1/2}$, and $G=PM_0^{-1}$.
Let $\beta(G) \equiv \beta({\bf G}_{\bot},G^+)\in SL(2,C)$ be the matrix
with the components
\begin{equation}
\beta_{11}=\beta_{22}^{-1}=2^{1/4}(G^+)^{1/2},\quad \beta_{12}=0,\quad
\beta_{21}=(G^x+\imath G^y)\beta_{22}
\label{13}
\end{equation}
and
\begin{equation}
k_i=L[\beta(G)]^{-1}p_i \quad (i=1,...A)
\label{14}
\end{equation}
The four-vectors $p_i$ have the components $(\omega({\bf p}_i),
{\bf p}_i)$, and the four-vectors $k_i$ have the components
$(\omega({\bf k}_i),{\bf k}_i)$ where $\omega({\bf k})=(m^2+{\bf
k}^2)^{1/2}$. In turn, only $A-1$ vectors ${\bf k}_i$ are
independent since, as follows from Eqs. (\ref{13}) and (\ref{14}),
${\bf k}_1+...+{\bf k}_A=0$. Therefore $L[\beta(G)]$ has the meaning
of the boost, and ${\bf k}_i$ are the momenta in the c.m. frame.
It is easy to show that $M_0=\omega({\bf k}_1)+...+\omega({\bf k}_A)$.

\begin{sloppypar}
 It follows from direct calculations that
\begin{eqnarray}
&&\prod_{i=1}^{A} d\rho({\bf p}_{i\bot},p_i^+)=d\rho({\bf G}_{\bot},G^+)
d\rho(int), \nonumber \\
&&d\rho(int)=2(2\pi)^3 M_0^3\delta^{(3)}({\bf k}_1+\cdots +{\bf k}_A)
 \prod_{i=1}^{A}d\rho({\bf k}_{i\bot},k_i^+)
\label{15}
\end{eqnarray}
We also define the "internal' space $H_{int}$ as the space of
functions $\chi({\bf k}_1,\sigma_1,...{\bf k}_A,\sigma_A)$
such that
\begin{equation}
||\chi||^2=\sum_{\sigma_1...\sigma_A}\int\nolimits
|\chi({\bf k}_1,\sigma_1,...{\bf k}_A,\sigma_A)|^2
d\rho(int)\quad < \infty
\label{16}
\end{equation}
Then the space of functions satisfying Eq. (\ref{11}) can be realized
as the space of functions $\varphi ({\bf G}_{\bot},G^+)$ with the range
in $H_{int}$ and such that
\begin{equation}
\int\nolimits ||\varphi({\bf G}_{\bot},G^+)||^2
d\rho({\bf G}_{\bot},G^+)\quad < \infty
\label{17}
\end{equation}
\end{sloppypar}

 For the system of $A$ interacting particles the four-velocity operator
${\hat G}$ is not generally speaking free, but is unitarily
equivalent to $G$ \cite{sok}: ${\hat G}=BGB^{-1}$ where $B$ is the
Sokolov packing operator. If $A=2$ then it is possible to choose $B=1$
(see the discussion in \cite{prd}), but for $A\geq 3$ the operator $B$
in the point form is necessarily nontrivial \cite{sok}. The mass
operator ${\hat M}$ in the general case acts not only through the
variables of the space $H_{int}$, but also through $G$, but the
operator $B$ can be chosen in such a way that the operator
${\tilde M}=B^{-1}{\hat M}B$ acts only through the variables of the space
$H_{int}$ \cite{sok,riv,lev}. If the nucleus is in the bound state then
its internal WF $\chi\in H_{int}$ is the eigenfunction of the operator
${\tilde M}$ with the eigenvalue $m_A$: ${\tilde M}\chi = m_A \chi$.

 Taking into account the normalization of free states in the scattering
theory, the WF of the free one-nucleon state with the four-momentum
$p"$ and the spin projection $\sigma"$ can be written in the form
\begin{equation}
|p",\sigma"\rangle =2(2\pi)^3p^{"+}\delta^{(2)}({\bf p}_{\bot}-
{\bf p}_{\bot}")\delta(p^+-p^{"+})\delta_{\sigma\sigma"}
\label{18}
\end{equation}
where $\delta_{\sigma\sigma"}$ is the Cronecker symbol.

 Analogously if the initial nucleus has the four-velocity $G'$, its WF
has the form \cite{riv,lev}
\begin{equation}
|G',\chi\rangle = B\frac{2}{m_A} (2\pi)^3G^{'+}
\delta^{(2)}({\bf G}_{\bot}-{\bf G}_{\bot}')\delta(G^+-G^{'+})\chi
\label{19}
\end{equation}
where the internal WF is normalized as $||\chi||=1$.

\section{General formulas for deep inelastic scattering}
\label{S3}

 It is well known that the DIS cross-section is defined by the tensor
\begin{equation}
W^{\mu\nu}=\frac{1}{4\pi}\sum_{X}(2\pi)^4\delta^{(4)}(P'+q-P_X)
\langle i|{\hat J}^{\mu}(0)|X\rangle
\langle X|{\hat J}^{\nu}(0)|i\rangle
\label{20}
\end{equation}
where a sum is taken over all possible final states $|X\rangle$,
and $P_X$ is the 4-momentum of the state $|X\rangle$. As follows from
relativistic invariance and current conservation, the tensor
$W^{\mu\nu}$ can be written in the form
\begin{eqnarray}
&&W^{\mu\nu}(P',q)=F_1(x,q^2)(\frac{q^{\mu}q^{\nu}}{q^2}-
\eta^{\mu\nu})+\nonumber\\
&&\frac{F_2(x,q^2)}{(P'q)}(P^{'\mu}-\frac{q^{\mu}(P'q)}{q^2})
(P^{'\nu}-\frac{q^{\nu}(P'q)}{q^2})+\nonumber\\
&&\frac{\imath g_1(x,q^2)}{(P'q)}e^{\mu\nu\rho\lambda}q_{\rho}
S_{\lambda}-\imath \frac{F_3(x,q^2)}
{2(P'q)}e^{\mu\nu\rho\lambda}P'_{\rho}q_{\lambda}
\label{21}
\end{eqnarray}
Here $\epsilon^{\mu\nu\rho\lambda}$ is the absolutely antisymmetric
tensor ($e^{0123}=1$) and $S_{\lambda}$ is the relativistic spin
operator. We do not write down terms which vanish in the Bjorken limit,
depend on the tensor polarization etc. The term with $F_3$ is present
only in the case of the neutrino (antineutrino) - nucleus scattering.

 Equation (\ref{21}) can also be written for DIS on the nucleon. In this
case we use ${\tilde W}^{\mu\nu}(p',{\tilde q})$ to denote the hadronic
tensor, $p'$ to denote the four-momentum of the initial nucleon,
${\tilde q}$ to denote the four-momentum of the virtual photon or
W-boson, and ${\tilde x}$ to denote the Bjorken variable
$-{\tilde q}^2/2p'{\tilde q}$.

 We shall always assume that the nucleus under consideration has the
lowest mass among all states with the baryon number equal to $A$. Then,
as follows from Eq. (\ref{8}), if $q^2<0$ then the structure functions
are not equal to zero only if $x\in (0,1]$. Analogously, in nucleon DIS
the structure functions are not equal to zero only if
\begin{equation}
\label{22}
{\tilde q}^2\frac{1-{\tilde x}}{{\tilde x}}\leq 0
\end{equation}
If ${\tilde q}^2<0$ then ${\tilde x}\in (0,1]$, but this condition is
also satisfied for ${\tilde q}^2>0$ if ${\tilde x}<0$ or ${\tilde x}>1$.

 Let us consider the nucleon DIS in the reference frame where $p^{'z}>0$,
${\bf p}_{\bot}={\tilde {\bf q}}_{\bot}=0$. Then, as follows from Eq.
(\ref{21}),
\begin{equation}
{\tilde W}^{--}=\frac{F_L^N({\tilde x},{\tilde q}^2){\tilde q}^-}
{4{\tilde x}^2p^{'+}}, \quad
{\tilde W}^{++}=\frac{F_L^N({\tilde x},{\tilde q}^2)p^{'+}}
{4{\tilde q}^-}
\label{23}
\end{equation}
where $F_L^N=F_2^N-2{\tilde x}F_1^N$ is the longitudinal nucleon
structure function (the index $N$ takes the values $p$ and $n$ for
the proton and neutron respectively). Another consequence of Eq.
(\ref{21}) is that if the initial nucleon is fully polarized along the
$z$ axis, then the tensor ${\tilde W}^{jl}$ for $j,l=1,2$ has the form
\begin{equation}
{\tilde W}_{\pm}^{jl}(p',{\tilde q})=F_1^N({\tilde x},{\tilde q}^2)
\delta_{jl}-\frac{\imath}{2}\epsilon_{jl}F_3^N({\tilde x},{\tilde q}^2)
\pm \imath \epsilon_{jl}g_1^N({\tilde x},{\tilde q}^2)
\label{24}
\end{equation}
where $\epsilon_{12}=-\epsilon_{21}=1$, $\epsilon_{11}=\epsilon_{22}=0$,
and $\pm$ corresponds to the $z$ projection of the spin equal to $\pm 1/2$.

\section{Deep inelastic scattering on the deuteron}
\label{S4}

 We assume that particle 1 in the deuteron is the proton and particle 2 is
the neutron. Then, as follows from Eqs. (\ref{15}) and (\ref{16}), the norm
of the internal deuteron WF is given by
\begin{equation}
||\chi||^2=\sum_{\sigma_p\sigma_n}\int\nolimits
|\chi({\bf k},\sigma_p,\sigma_n)|^2\frac{M_0({\bf k})^3d^3{\bf k}}
{2(2\pi)^3\omega({\bf k})^2}
\label{25}
\end{equation}
where ${\bf k}\equiv {\bf k}_1$ and $M_0({\bf k})=2\omega({\bf k})$.
As follows from the minimal relativity principle
\cite{BrKuo,CoPi,Kon,Coes} (see also the discussion in Ref. \cite{riv}),
the relativistic WF $\chi$ can be taken as in the standard nuclear
physics. Therefore we can take $\chi$ in the form
\begin{equation}
\chi({\bf k},\sigma_p,\sigma_n)=\frac{1}{\sqrt{2}M_0({\bf k})}
[\varphi_0(k)\delta_{jl}-
\frac{1}{\sqrt{2}}(\delta_{jl}-3\frac{k_jk_l}{k^2})\varphi_2(k)]
e_l(\tau_j\tau_2)_{\sigma_p\sigma_n}
\label{26}
\end{equation}
where $k=|{\bf k}|$, ${\bf e}$ is the polarization vector of the deuteron,
$\tau_i$ are the Pauli matrices and a sum over repeated indices $j,l=1,2,3$
is assumed. The functions $\varphi_0(k)$ and $\varphi_2(k)$ are the
amplitudes of the $S$ and $D$ states in the deuteron, and the normalization
is chosen such that
\begin{equation}
\int\nolimits[\varphi_0(k)^2+\varphi_2(k)^2]
\frac{d^3{\bf k}}{(2\pi)^3\omega({\bf k})}= 1
\label{27}
\end{equation}

 As noted in Sec. \ref{S2}, one can choose $B=1$ in the deuteron case and
therefore the IA is compatible with the continuity equation (see Sec.
\ref{S1}). The calculations in the parton model and in the model where
the IA is used in the point form \cite{Web,hep} show that if the IA is
valid then the constituents comprising the system under consideration
interact with the virtual photon incoherently. Therefore it is reasonable
to assume that the cross-section of the DIS on the deuteron is a sum of
the cross-sections corresponding to the DIS on the proton and neutron. It
is also reasonable to assume that the final state interaction between
the spectator nucleon and the particles produced in the DIS on the
struck nucleon can be neglected if $|q^2|$ is very large.

 With all these assumptions we can write Eq. (\ref{20}) in the form
\begin{eqnarray}
&&W^{\mu\nu}(P',q)=\frac{1}{4\pi}\sum_{\sigma_n}\sum_{{\tilde X}}
\int\nolimits (2\pi)^4\delta^{(4)}(P'+q-p_n-P_{\tilde X})
d\rho({\bf p}_{n\bot},p_n^+)\cdot\nonumber\\
&&\langle G'\chi|J^{\mu}_p(0)|p_n,\sigma_n;{\tilde X}\rangle
\langle p_n,\sigma_n;{\tilde X}|J^{\nu}_p(0)|G'\chi\rangle + (...)
\label{28}
\end{eqnarray}
and one should bear in mind that this expression is valid for those
$\mu,\nu$ for which the deuteron CO can be taken in the IA.
Here ${\tilde X}$ are all possible final states in the DIS on the proton,
$P_{\tilde X}$ is the four-momentum of the state ${\tilde X}$,
$J^{\mu}_p(0)$ is the proton CO, the state
$|p_n,\sigma_n;{\tilde X}\rangle$ is the tensor product of the states
$|p_n,\sigma_n\rangle$ and $|{\tilde X}\rangle$
and (...) is the contribution from the
DIS on the neutron which can be written analogously.

 Let us introduce the function
\begin{equation}
\Phi(p_p,p_n,\sigma_p,\sigma_n) = \frac{2}{m_d} (2\pi)^3G^{'+}
\delta^{(2)}({\bf G}_{\bot}-{\bf G}_{\bot}')\delta(G^+-G^{'+})
\chi({\bf k},\sigma_p,\sigma_n)
\label{29}
\end{equation}
where $m_d$ is the mass of the deuteron. Then as follows from Eqs.
(\ref{12}), (\ref{18}) and (\ref{19})
\begin{equation}
|G',\chi\rangle = \sum_{\sigma_p\sigma_n}\int\nolimits
\Phi(p_p,p_n,\sigma_p,\sigma_n) |p_p,\sigma_p\rangle |p_n,\sigma_n\rangle
d\rho({\bf p}_{p\bot},p_p^+)d\rho({\bf p}_{n\bot},p_n^+)
\label{30}
\end{equation}

 As follows from Eqs. (\ref{18}) and (\ref{30}), Eq. (\ref{28}) can be
written in the form
\begin{eqnarray}
&&W^{\mu\nu}(P',q)=\frac{1}{4\pi}\sum_{\sigma_p\sigma_p'\sigma_n}
\sum_{{\tilde X}}\int\nolimits (2\pi)^4\delta^{(4)}(P'+q-
p_n-P_{\tilde X}) \cdot\nonumber\\
&&\Phi(p_p,p_n,\sigma_p,\sigma_n)^*\Phi(p_p',p_n,\sigma_p',\sigma_n)
\langle p_p,\sigma_p|J^{\mu}_p(0)|{\tilde X}\rangle
\langle {\tilde X}|J^{\nu}_p(0)|p_p',\sigma_p'\rangle\cdot\nonumber\\
&&d\rho({\bf p}_{p\bot},p_p^+)d\rho({\bf p}_{p\bot}',p_p^{'+})
d\rho({\bf p}_{n\bot},p_n^+)+ (...)
\label{31}
\end{eqnarray}

 Now we use the following relation which can be verified by a direct
calculation \cite{sok}: if $p_n$ is fixed and $G=(p_p+p_n)/|p_p+p_n|$,
where the modulus of the four-vector is understood in the Lorentz
metric, then
\begin{equation}
d\rho({\bf p}_{p\bot},p_p^+)=\frac{|p_p+p_n|^4}{(p_p,p_p+p_n)}
d\rho({\bf G}_{\bot},G^+)
\label{32}
\end{equation}
Since $|p_p+p_n|=M_0({\bf k})$, $(p_p,p_p+p_n)=\omega({\bf k})
M_0({\bf k})$, it follows from Eqs. (\ref{29}), (\ref{31}) and
(\ref{32}) that
\begin{eqnarray}
&&W^{\mu\nu}(P',q)=\frac{1}{4\pi}\sum_{\sigma_p\sigma_p'\sigma_n}
\sum_{{\tilde X}}\int\nolimits (2\pi)^4\delta^{(4)}(P'+q-
p_n-P_{\tilde X}) \cdot\nonumber\\
&&\frac{M_0({\bf k})^6}{m_d^2\omega({\bf k})^2}
\chi({\bf k},\sigma_p,\sigma_n)^*\chi({\bf k},\sigma_p',\sigma_n)
\langle p_p,\sigma_p|J^{\mu}_p(0)|{\tilde X}\rangle \cdot\nonumber\\
&&\langle {\tilde X}|J^{\nu}_p(0)|p_p,\sigma_p'\rangle
d\rho({\bf p}_{n\bot},p_n^+)+ (...)
\label{33}
\end{eqnarray}
where the four-vector $p_p$ is a function of $p_n$ and $G'$ defined by
the condition $(p_p+p_n)/|p_p+p_n|=G'$, and ${\bf k}$ is the spatial part
of the four-vector $k_1$ defined by Eq. (\ref{14}) with $G=G'$.

 The next step is to change the variables from ${\bf p}_{n\bot},p_n^+$ to
${\bf k}$. For this purpose we can use Eq. (\ref{12}) and write
\begin{equation}
d\rho({\bf p}_{n\bot},p_n^+)=d\rho({\bf p}_{n\bot},p_n^+)
\int\nolimits 2(2\pi)^3p_p^+\delta^{(2)}({\bf p}_{\bot}-{\bf p}_{\bot}')
\delta(p^+-p^{'+})d\rho({\bf p}_{p\bot},p_p^+)
\label{34}
\end{equation}
Then using Eqs. (\ref{15}) and (\ref{32}) we get
\begin{equation}
d\rho({\bf p}_{n\bot},p_n^+)=\frac{\omega({\bf k})d\rho(int)}
{M_0({\bf k})^3}
\label{35}
\end{equation}
Therefore using again Eq. (\ref{15}) we can write Eq. (\ref{33}) in the
form
\begin{eqnarray}
&&W^{\mu\nu}(P',q)=\frac{1}{4\pi}\sum_{\sigma_p\sigma_p'\sigma_n}
\sum_{{\tilde X}}\int\nolimits (2\pi)^4\delta^{(4)}(P'+q-
p_n-P_{\tilde X}) \cdot\nonumber\\
&&\frac{M_0({\bf k})^3}{m_d^2\omega({\bf k})}
\chi({\bf k},\sigma_p,\sigma_n)^*\chi({\bf k},\sigma_p',\sigma_n)
\langle p_p,\sigma_p|J^{\mu}_p(0)|{\tilde X}\rangle \cdot\nonumber\\
&&\langle {\tilde X}|J^{\nu}_p(0)|p_p,\sigma_p'\rangle
d\rho(int)+ (...)
\label{36}
\end{eqnarray}
where $p_p$ and $p_n$ are the four-vectors depending on $G'$ and
${\bf k}$ as (see Eq. (\ref{14}))
\begin{equation}
p_p=L[\beta(G)]k_p \quad p_n=L[\beta(G)]k_n
\label{37}
\end{equation}
where $k_p=(\omega({\bf k}),{\bf k})$, $k_n=(\omega({\bf k}),-{\bf k})$.

 The purpose of exposing the detailed derivation of Eq. (\ref{36})
from Eq. (\ref{28}) is to stress that this derivation does not use any
additional assumptions, i.e. Eq. (\ref{36}) is an unambiguous consequence
of Eq. (\ref{28}).

 Let us suppose that the deuteron is fully polarized along the
positive direction of the $z$ axis. In this case
$e_je_l^*=(\delta_{jl}/3)-(\imath \epsilon_{jl}/2)$ where we do not
include the contribution of the tensor polarization. Then a direct
calculation using Eqs. (\ref{15}), (\ref{20}), (\ref{26}) and (\ref{36})
gives
\begin{eqnarray}
&&W^{\mu\nu}(P',q)=\int\nolimits \{[\varphi_0(k)^2+\varphi_2(k)^2]
[{\tilde W}_{+p}^{\mu\nu}(p_p,{\tilde q})+
{\tilde W}_{-p}^{\mu\nu}(p_p,{\tilde q})]+\nonumber\\
&&[\varphi_0(k)-\frac{\varphi_2(k)}{\sqrt{2}}][\varphi_0(k)+
\frac{\varphi_2(k)}{\sqrt{2}}(\frac{3{\bf k}_{\bot}^2}{k^2}-1)]
[{\tilde W}_{+p}^{\mu\nu}(p_p,{\tilde q})-
{\tilde W}_{-p}^{\mu\nu}(p_p,{\tilde q})]\}\cdot\nonumber\\
&&\frac{M_0({\bf k})^3d^3{\bf k}}{2(2\pi)^3\omega({\bf k})^2m_d^2}+(...)
\label{38}
\end{eqnarray}
where ${\tilde q}=q-[M_0({\bf k})-m_d]G'$, and therefore, as follows
from Eq. (\ref{9}), the four-vector ${\tilde q}$ has the following
components in the reference frame considered in Sec. \ref{S1}:
\begin{equation}
{\tilde q}^0=(M"-M_0({\bf k}))G^{'0},\quad
{\tilde {\bf q}}=-(M"+M_0({\bf k})){\bf G}'
\label{39}
\end{equation}

 It is clear from Eq. (\ref{38}) that the nucleon structure functions
entering into this expression depend on
${\tilde x}=-{\tilde q}^2/2(p_p{\tilde q})$. As follows from Eqs.
(\ref{37}) and (\ref{39}), in the Bjorken limit
\begin{equation}
{\tilde x}=\frac{M_0({\bf k})-m_d(1-x)}{\omega({\bf k})+k^z}, \quad
{\tilde q}^2=-\frac{|q^2|}{m_dx}[M_0({\bf k})-m_d(1-x)]
\label{40}
\end{equation}
If $x>1$ then it follows from Eq. (\ref{40}) that ${\tilde q}^2<0$,
${\tilde x}>1$. If $x<0$ and $M_0({\bf k})-m_d(1-x)>0$ then
${\tilde q}^2>0$, ${\tilde x}>0$. Finally, if $x<0$ and
$M_0({\bf k})-m_d(1-x)<0$ then ${\tilde q}^2<0$, ${\tilde x}<0$. In all
these cases Eq. (\ref{22}) is not satisfied and therefore the nucleon
structure functions are equal to zero. We conclude that the deuteron
structure functions automatically satisfy the condition that they are
equal to zero if $x<0$ and $x>1$, as it should be.

 Let us introduce the internal variable
$\xi=[\omega({\bf k})+k^z]/M_0({\bf k})$. Then it is easy to show that
$\xi \in (0,1)$ and
\begin{eqnarray}
&&M_0({\bf k})\equiv M_0({\bf k}_{\bot},\xi)=\frac{m^2+{\bf k}_{\bot}^2}
{\xi(1-\xi)},\quad k^z=(\xi-\frac{1}{2})M_0({\bf k}_{\bot},\xi), \nonumber\\
&&\frac{d^3{\bf k}}{\omega({\bf k})}=\frac{d^2{\bf k}_{\bot}d\xi}
{2\xi(1-\xi)}
\label{41}
\end{eqnarray}
Therefore the sets $({\bf k}_{\bot},k^z)$ and $({\bf k}_{\bot},\xi)$
have a one-to-one relation to each other, and the normalization
condition (\ref{27}) can be written in the form
\begin{equation}
\int\nolimits d^2{\bf k}_{\bot}\int_{0}^{1}
\frac{d\xi}{2(2\pi)^3\xi(1-\xi)}[\varphi_0(k)^2+\varphi_2(k)^2]=1
\label{42}
\end{equation}
where $k=({\bf k}_{\bot}^2+k_z^2)^{1/2}$ is the function of
${\bf k}_{\bot},\xi$. However not all $\xi's$ contribute to Eq. (\ref{38})
since, as follows from Eq. (\ref{40}), ${\tilde x}\leq 1$ only if
$\xi \in [\xi_{min},1]$ where
\begin{equation}
\xi_{min}\equiv \xi_{min}({\bf k}_{\bot},x)=\frac{m^2+{\bf k}_{\bot}^2}
{m^2+{\bf k}_{\bot}^2+m_d^2(1-x)^2}
\label{43}
\end{equation}

 Now everything is ready to write down the explicit expressions for the
deuteron structure functions in terms of the nucleon ones. Since
$p_p^+\gg |{\bf k}_{\bot}|$ in the Bjorken limit, we can use Eq.
(\ref{24}). Let us recall that Eq. (\ref{38}) is valid for $\mu,\nu=1,2$
since the $\bot$ components of the CO are not constrained by the
continuity equation. Therefore, as follows from Eqs. (\ref{24}),
(\ref{38}) and (\ref{41})
\begin{eqnarray}
&&F_i^d(x,q^2)= \int\nolimits d^2{\bf k}_{\bot}\int_{\xi_{min}}^{1}
\frac{d\xi}{(2\pi)^3\xi(1-\xi)}[\frac{M_0({\bf k}_{\bot},\xi)}
{m_d}]^2\cdot\nonumber\\
&&[\varphi_0(k)^2+\varphi_2(k)^2][F_i^p({\tilde x},{\tilde q}^2)+
F_i^n({\tilde x},{\tilde q}^2)],\quad (i=1,3)
\label{44}
\end{eqnarray}
\begin{eqnarray}
&&g_1^d(x,q^2)= \int\nolimits d^2{\bf k}_{\bot}\int_{\xi_{min}}^{1}
\frac{d\xi}{(2\pi)^3\xi(1-\xi)}[\frac{M_0({\bf k}_{\bot},\xi)}
{m_d}]^2[\varphi_0(k)-\frac{\varphi_2(k)}{\sqrt{2}}]\cdot\nonumber\\
&&[\varphi_0(k)+\frac{\varphi_2(k)}{\sqrt{2}}(\frac{3{\bf k}_{\bot}^2}
{k^2}-1)][g_1^p({\tilde x},{\tilde q}^2)+g_1^n({\tilde x},
{\tilde q}^2)]
\label{45}
\end{eqnarray}
where ${\tilde x}$ and ${\tilde q}^2$ are given by Eq. (\ref{40}) and
$M_0({\bf k})$ and $k$ are the functions of ${\bf k}_{\bot},\xi$ defined
by Eq. (\ref{41}).

 Let us now consider the longitudinal structure function. If the minus
component of the CO is taken in the IA then, as follows from Eqs.
(\ref{21}), (\ref{23}), (\ref{37}), (\ref{38}) and (\ref{41})
\begin{eqnarray}
&&F_L^d(x,q^2)= \int\nolimits d^2{\bf k}_{\bot}\int_{\xi_{min}}^{1}
\frac{d\xi}{(2\pi)^3\xi^2(1-\xi)}[\frac{M_0({\bf k}_{\bot},\xi)}
{m_d}](\frac{x}{{\tilde x}})^2\cdot\nonumber\\
&&[\varphi_0(k)^2+\varphi_2(k)^2][F_L^p({\tilde x},{\tilde q}^2)+
F_L^n({\tilde x},{\tilde q}^2)]
\label{46}
\end{eqnarray}
Here we have taken into account that, as follows from Eqs. (\ref{9})
and (\ref{39}), $q^-={\tilde q}^-$ in the Bjorken limit. An analogous
derivation in the case when the plus component of the CO is taken in
the IA gives
\begin{eqnarray}
&&F_L^d(x,q^2)= \int\nolimits d^2{\bf k}_{\bot}\int_{\xi_{min}}^{1}
\frac{d\xi}{(2\pi)^3(1-\xi)}[\frac{M_0({\bf k}_{\bot},\xi)}
{m_d}]^3\cdot\nonumber\\
&&[\varphi_0(k)^2+\varphi_2(k)^2][F_L^p({\tilde x},{\tilde q}^2)+
F_L^n({\tilde x},{\tilde q}^2)]
\label{47}
\end{eqnarray}

\section{Structure functions of heavier nuclei}
\label{S5}

 If the IA is valid, the nucleons absorb the virtual photon incoherently
and the final state interaction between the spectator nucleons and the
particles produced in the DIS on the struck nucleon can be neglected then
as follows from Eq. (\ref{20})
\begin{eqnarray}
&&W^{\mu\nu}(P',q)=\frac{1}{4\pi}\sum_{i=1}^{A}\sum_{Y_i,{\tilde X}_i}
(2\pi)^4\delta^{(4)}(P'+q-P_{Y_i}-P_{{\tilde X}_i})\cdot\nonumber\\
&&\langle G'\chi|J^{\mu}_i(0)|Y_i;{\tilde X}_i\rangle
\langle Y_i;{\tilde X}_i|J^{\nu}_i(0)|G'\chi\rangle
\label{48}
\end{eqnarray}
Here ${\tilde X}_i$ are all possible final states in the DIS on the $i$-th
nucleon, $Y_i$ are all possible states of the final system of the
nucleons $1,...i-1,i+1,...A$, $P_{{\tilde X}_i}$ and $P_{Y_i}$ are the
corresponding four-momenta and $J^{\mu}_i(0)$ is the CO of the $i$-th
nucleon.

 The calculation of the nucleus structure functions from Eq. (\ref{48})
is much more complicated than the calculation of the deuteron structure
functions from Eq. (\ref{28}). The reasons of the complications are as
follows. Firstly, as noted in Sec. \ref{S2}, the operator $B$ is not
equal to unity if $A\geq 3$ (the explicit expressions for $B$ were
written only for $A=3$  though in principle this expression can be
written for any $A$ \cite{sok,riv}). Secondly, $P_{Y_i}$ is not equal to
a sum of the four-momenta of the nucleons $1,...i-1,i+1,...A$ since
there is no ground to neglect the final state interaction in the
system $(1,...i-1,i+1,...A)$ and for the same reason $Y_i$ is not an
antisymmetrized tensor product of the one-nucleon states. Finally, the
relativistic WF $\chi$ has never been calculated.

 If we assume that the nucleus is the nonrelativistic system with a
good accuracy then we can take for $\chi$ its nonrelativistic expression
and neglect the operator $B-1$. Let us also assume that in the
nonrelativistic approximation the final state interaction in the
system $(1,...i-1,i+1,...A)$ can be neglected. Once these assumptions
are made, the nucleus structure functions can be calculated from Eq.
(\ref{48}) by analogy with the calculation of the deuteron structure
functions in the preceding section.

 The result of the calculation is as follows. We introduce the
quantities ${\tilde q}=q-(M_0-m_A)G'$ and
${\tilde x}=-{\tilde q}^2/2(p_1{\tilde q})$. Then in the Bjorken limit
\begin{equation}
{\tilde x}=\frac{M_0-m_A(1-x)}{\omega_1({\bf k}_1)+k_1^z}, \quad
{\tilde q}^2=-\frac{|q^2|}{m_Ax}[M_0-m_A(1-x)]
\label{49}
\end{equation}
If $Z$ is the number of protons in the nucleus then
\begin{eqnarray}
&&F_i^A(x,q^2)=\sum_{\sigma_1...\sigma_A}\int\nolimits
\frac{M_0^3}{m_A^2\omega_1({\bf k}_1)}[ZF_i^p({\tilde x},{\tilde q}^2)+
(A-Z)F_i^n({\tilde x},{\tilde q}^2)]\cdot\nonumber\\
&&|\chi({\bf k}_1,\sigma_1,...{\bf k}_A,\sigma_A)|^2
d\rho(int)\quad (i=1,3)
\label{50}
\end{eqnarray}
and, if the minus component of the CO is taken in the IA then
\begin{eqnarray}
&&F_L^A(x,q^2)=\sum_{\sigma_1...\sigma_A}\int\nolimits
\frac{M_0^3}{m_A\omega_1({\bf k}_1)[\omega_1({\bf k}_1)+k_1^z]}
(\frac{x}{\tilde x})^2[ZF_L^p({\tilde x},{\tilde q}^2)+\nonumber\\
&&(A-Z)F_L^n({\tilde x},{\tilde q}^2)] |\chi({\bf k}_1,
\sigma_1,...{\bf k}_A,\sigma_A)|^2d\rho(int)
\label{51}
\end{eqnarray}
It is also possible to obtain the expressions for the spin dependent
structure functions but we shall not dwell on this problem.

 If we introduce the quantity $\xi=[\omega_1({\bf k}_1)+k_1^z]/M_0$
then it is easy to show that the integration over $\xi$ in Eqs.
(\ref{50}) and (\ref{51}) is in fact performed not from 0 to 1 but
from $\xi_{min}$ to 1 where
\begin{eqnarray}
&&\xi_{min}\equiv \xi_{min}({\bf k}_{1\bot},M_1,x)=
\frac{1}{2}\{1-\alpha-\beta+[(1-\alpha-\beta)^2+
4\alpha]^{1/2}\}, \nonumber\\
&&\alpha=\frac{m^2+{\bf k}_{1\bot}^2}{M_1^2-m^2},\quad
\beta=\frac{m_A^2(1-x)^2}{M_1^2-m^2}
\label{52}
\end{eqnarray}
and $M_1$ is the free mass of the system $(2,...A)$.

 The validity of the assumptions used in the derivation of Eqs.
(\ref{50}) and (\ref{51}) is questionable. Indeed, we neglected the
quantities $B-1$ and $p_1+...p_{i-1}+p_{i+1}+...p_A-P_{Y_i}$ since
they are small in the nonrelativistic approximation. However in this
approximation it is also possible to neglect the difference between
$M_0$ and $m_A$. We shall discuss this question in the next section.

\section{Discussion}
\label{S6}

 It is usually believed that DIS with very large momentum transfer can
be described in the IA. As shown in Sec. \ref{S1}, this approximation is
compatible with Lorentz invariance only in the point form of dynamics.
Then, assuming that the deuteron is the relativistic system of the proton
and neutron, it is possible to obtain the unambiguous expressions for the
deuteron structure functions $F_1^d(x,q^2)$, $F_3^d(x,q^2)$ and
$g_1^d(x,q^2)$ (see Eqs. (\ref{44}) and (\ref{45})) in terms of the
corresponding nucleon structure functions. At the same time, the result
for the deuteron longitudinal structure function depends on which component
of the CO in the $+-$ plane is taken in the IA.

 The main difference between our result and the results of many authors
(see, for example, Refs. \cite{Ak,FS,Bick,COS,Ciofi,MelKul}) is that in our
approach the nucleon structure functions entering into the integrals
defining the corresponding deuteron structure functions depend on
${\tilde x}$ and ${\tilde q}^2$ given by Eq. (\ref{40}) while the
standard expressions do not contain terms with $M_0({\bf k})-m_d$:
\begin{equation}
{\tilde x}=\frac{m_dx}{\omega({\bf k})+k^z}, \quad
{\tilde q}^2=q^2
\label{53}
\end{equation}
These expressions have the well-known partonic interpretation since in
this interpretation the effect of binding is considered as a higher
twist effect. In particular, the expression for ${\tilde x}$ is
compatible with the flux factor in Refs.
\cite{Ak,FS,Bick,COS,Ciofi,MelKul}). Equation (\ref{53}) can be also
obtained in the front form of dynamics \cite{COS,Web,hep}, but, as noted
in Sec. \ref{S1}, the CO in
this form does not properly commute with the Lorentz boost generators.
Let us stress that in our approach the particles are always on-shell
and the nucleon structure functions in Eqs. (\ref{44}-{47}) refer only
to real nucleons. Therefore the difference between Eqs. ({40}) and
(\ref{53}) can be regarded as an analog of off-shellness in our approach.

 The experimental data are usually given in terms of not the Bjorken
variable $x$ but in terms of $x_{eff}=m_Ax/m$. Therefore the results
(\ref{40}) and (\ref{53}) are practically the same when $mx_{eff}\gg
M_0({\bf k})-m_d$ and considerably differ each other when
$mx_{eff}
\hbox{{\normalsize\,\,%
\lower -.6ex \hbox{$<$}\kern -.75em \lower .5ex \hbox{$\sim$}\,\,}}
M_0({\bf k})-m_d$. The quantity $M_0({\bf k})-m_d$ can
be written as $T+|\epsilon_d|$ where $T$ is the kinetic energy of the
nucleons in the deuteron and $\epsilon_d=-2.23MeV$ is the deuteron
binding energy. According to the well-known phenomenological models
of the nucleon-nucleon potential, the average value of $T$ is of order
$20MeV$. Therefore our result coincides with the standard one at
$x_{eff}\gg 0.02$ and considerably differs from it at $x_{eff}
\hbox{{\normalsize\,\,%
\lower -.6ex \hbox{$<$}\kern -.75em \lower .5ex \hbox{$\sim$}\,\,}}
 0.02$.

 If $0.02\ll x_{eff}<1$ and $x_{eff}$ is not too close to 1, the main
contribution to the deuteron structure functions is given by the
nonrelativistic region of the deuteron WF where $|{\bf k}|\ll m$. Then, as
easily follows from Eqs. (\ref{42}-{47})
\begin{eqnarray}
&&F_i^d(x,q^2)=2[F_i^p(x_{eff},q^2)+F_i^n(x_{eff},q^2)]\quad
(i=1,3),\nonumber\\
&&g_1^d(x,q^2)=2[g_1^p(x_{eff},q^2)+g_1^n(x_{eff},q^2)]
(1-\frac{3}{2}P_D),\nonumber\\
&&F_L^d(x,q^2)=F_L^p(x_{eff},q^2)+F_L^n(x_{eff},q^2)
\label{54}
\end{eqnarray}
where $x_{eff}=2x$ and $P_D$ is the probability of the $D$ state in the
deuteron. In this case the result for $F_L^d$ does not depend on the
choice of Eq. (\ref{46}) or ({\ref{47}). As a consequence of Eq.
(\ref{54}), $F_2^d(x,q^2)=F_L^p(x_{eff},q^2)+F_2^n(x_{eff},q^2)$ what
is the well-known result in the case when nuclear effects in the
deuteron are neglected. Let us also note that if the main contribution
to the first moments $\Gamma_1^d$, $\Gamma_1^p$and $\Gamma_1^p$ of the
structure function $g_1(x)$ for the deuteron, proton and neutron
respectively is
given by $0.01\ll x<1$ and $0.02\ll x_{eff}<1$ then we obtain from Eq.
(\ref{54}) the well-known relation $\Gamma_1^d=(\Gamma_1^p+\Gamma_1^p)
(1-1.5P_D)$ used for extracting $\Gamma_1^n$ from the proton and
neutron data to test the Bjorken sum rule \cite{Bjor2}. The results
(\ref{54}) are well-known and, as shown by several authors (see, for
example, Ref. \cite{MST}), the nuclear corrections to these expressions
are small.

 Let us now consider the deuteron structure functions at $x
\hbox{{\normalsize\,\,%
\lower -.6ex \hbox{$<$}\kern -.75em \lower .5ex \hbox{$\sim$}\,\,}}
0.01$.
In the literature the role of shadowing at such values of $x$ is
widely discussed \cite{Br,BK,NZ,MelKul}.  One of the problems is the
behavior of the shadowing contribution at very large $|q^2|$. The
authors of Refs. \cite{Br,BK,MelKul} consider the scaling part of
this contribution (which does not depend on $|q^2|$) but, as noted in
Ref. \cite{NZ}, the shadowing contribution at large $|q^2|$ is a
logarythmically decreasing function. Therefore one might expect that
at large $|q^2|$ the IA is dominant (on the other hand, in the
electroproduction $|q^2|$ should be much smaller than $m_Z^2$).
Anyway, the shadowing contribution does not depend on the nucleon
structure functions. Therefore if we wish to extract the neutron
structure functions from the proton and deuteron data we have to know
the contribution of the IA. Meanwhile, as follows from Eq. (\ref{40}),
even if $x\ll 0.01$ the quantity ${\tilde x}$ is of order 0.02. The
minimum of ${\tilde x}$ is approximately equal to 0.002 at $x=0$,
${\bf k}_{\bot}=0$ and $k^z=|\epsilon_d|/2m\approx 1Mev/c$, but the
contribution of very small ${\bf k}$'s to the integrals (\ref{44}-{47})
is also small. We conclude that {\it there is no way to extract the
neutron structure functions  from the proton and deuteron data at
${\tilde x}
\hbox{{\normalsize\,\,%
\lower -.6ex \hbox{$<$}\kern -.75em \lower .5ex \hbox{$\sim$}\,\,}}
0.01$}.

 The problem of extracting the neutron structure functions from the
proton and deuteron data was considered by many authors (see, for
example, Ref. \cite{Th} and references cited therein). The authors of
Ref. \cite{Lead} argue that the Gottfried sum \cite{Got} is rather
sensitive to the pion admixture in the deuteron while the Bjorken sum
\cite{Bjor2} is not sensitive to such an admixture. Our conclusion
implies that even in an idealized case when the contribution of the
IA is known exactly, the Bjorken and Gottfried sum rules cannot be
tested. If the contribution of the region ${\tilde x}
\hbox{{\normalsize\,\,%
\lower -.6ex \hbox{$<$}\kern -.75em \lower .5ex \hbox{$\sim$}\,\,}}
0.01$ to
these sum rules is small then our conclusion is of only academic
interest. However the data of the SMC Collaboration on $g_1^p({\tilde x})$
\cite{RVD} and the dramatic rise of $F_2^p({\tilde x})$ discovered
recently at HERA \cite{H1} give grounds to think that the role of
${\tilde x}
\hbox{{\normalsize\,\,%
\lower -.6ex \hbox{$<$}\kern -.75em \lower .5ex \hbox{$\sim$}\,\,}}
0.01$ in the Bjorken and Gottfried sum rules is not
negligible.

 Let us now consider the results of Sec. \ref{S5} for the structure
functions of heavier nuclei. It is clear from Eq. (\ref{49}) that
these results agree with the standard ones if $mx_{eff}\gg M_0-m_A$.
We can write $M_0-m_A=A(T+|\epsilon|)$ where $T$ is the kinetic
energy per nucleon and $\epsilon$ is the binding per nucleon. For
heavy nuclei $\epsilon\approx 8MeV$, and the reasonable value of $T$ is
20MeV. Therefore $(M_0-m_A)/m$ becomes larger than unity at $A
\hbox{{\normalsize\,\,%
\lower -.6ex \hbox{$>$}\kern -.75em \lower .5ex \hbox{$\sim$}\,\,}}
40$.
We see that though $(M_0-m_A)/m$ is the term of order $v^2/c^2$, this
quantity cannot be neglected for heavy nuclei. Moreover, since we
cannot neglect $M_0-m_A$, we also cannot neglect the operator $B-1$
and the final state interaction in the system $(1,...i-1,i+1,...A)$. We
conclude that while the effect of the final state interaction is usually
believed to be a higher twist effect, in our approach this effect is
important even in the Bjorken limit.

 The above discussion shows that only the results of Sec. \ref{S4} for
the deuteron are reliable. Moreover, in our opinion the results for
the deuteron structure functions in terms of the proton and neutron
structure functions are much more reliable than the standard results for
the nucleon structure functions in terms of the quark structure
functions. The matter is that in the deuteron case the assumption that
the system under consideration is the relativistic system with a fixed
number of particles is reasonable, and it is reasonable to think that
the final state interaction can be neglected in the Bjorken limit.

 If the IA is dominant at high $|q^2|$ and small $x$ then the results
of Sec. \ref{4} make it possible to give the following predictions on
the behavior of the deuteron structure functions at $x\ll 0.01$. The
functions $F_1^d(x,q^2)$, $F_3^d(x,q^2)$ and $g_1^d(x,q^2)$ are given by
Eqs. (\ref{44}), and (\ref{45}) where, as follows from Eqs. (\ref{40}),
and (\ref{43}),
\begin{equation}
{\tilde x}=\frac{2\omega({\bf k})-m_d}{\omega({\bf k})+k^z}, \quad
{\tilde q}^2=-\frac{|q^2|}{m_dx}[2\omega({\bf k})-m_d],\quad
\xi_{min}=\frac{m^2+{\bf k}_{\bot}^2}
{m^2+{\bf k}_{\bot}^2+m_d^2}
\label{55}
\end{equation}
Therefore the $x$ dependence of these structure
functions is fully determined by the ${\tilde q}$ dependence of the
nucleon structure functions while ${\tilde x}$ does not depend on $x$
at $x \ll 0.01$. In
particular, if the ${\tilde q}$ dependence of the nucleon structure
functions is weak (e.g. logarythmic), the $x$ dependence of these
deuteron structure functions also is weak. At the same time, if the
longitudinal structure function is given by Eq. (\ref{46}) then the
ratio $R$ of the total cross sections for longitudinally and transversely
polarized virtual photons falls off as $\sim x^2$ and $F_2^d(x,q^2)$ falls
off as $\sim x$, while if the longitudinal structure function is given by
Eq. (\ref{47}) (what is less probable) then $R$ increases with the
decrease of $x$ as $\sim 1/x$ and the dependence $F_2^d(x,q^2)$ on $x$ is
weak.

 The most unusual feature of Eq. (\ref{55}) is that ${\tilde q}^2$
strongly differs from $q^2$ while in the covariant approach based on
Feynman diagrams (see the discussion in Sec. \ref{S1})
${\tilde q}^2$ is always equal to $q^2$.

 At present the low $x$ region has been investigated by the NMC and E665
Collaborations \cite{NMC} but the values of $|q^2|$ in these experiments
where small at small $x$. However the measurements of the deuteron
structure functions at small $x$ and large $|q^2|$ are planned at HERA.

 We see that the decrease of $F_2^d(x,q^2)$ with the decrease of $x$ may
be not only a consequence of shadowing but also a consequence of the
IA. In any case one might think that in view of the results obtained at
HERA \cite{H1} the described behavior of the deuteron structure functions
at low $x$ fully differs from the behavior of the nucleon structure
functions at such $x$. A possible reason is as follows. The central point
of the above description of the deuteron structure functions is the
relation between $m_dx$ and $M_0({\bf k})-m_d$. In nuclear physics the
difference between free mass operator and the mass of the bound system is
always positive. However in quark models this difference may be negative if
confinement is the main interaction between quarks. In this case the
above approach does not work, since it is obvious that if confinement is
taken into account then the final state interaction cannot be neglected.
The importance of taking into account the role of confinement in nucleon
DIS was pointed out by several authors (see the recent paper \cite{Gurv}
and references therein) but it is usually believed that confinement is
only a higher twist effect. However if the CO is taken in the point
form then confinement may play an important role even in the Bjorken
limit. We suppose to investigate this problem in future publications.

\vskip 0.2em
\begin{center} {\bf Acknowledgments} \end{center}
\vskip 0.2em

\begin{sloppypar}
 The author is grateful to R. Van Dantzig, S.Gevorkyan, M.Karliner,
E.Pace and G.Salme for valuable discussions. This work was supported
by grant No. 93-02-3754 from the Russian Foundation for Fundamental
Research.
\end{sloppypar}

\end{document}